\documentclass[a4paper,twoside]{article}

\usepackage{epsfig}
\usepackage{epstopdf}
\usepackage{subfigure}
\usepackage{calc}
\usepackage{amssymb}
\usepackage{amstext}
\usepackage{amsmath}
\usepackage{amsthm}
\usepackage{multicol}
\usepackage{pslatex}
\usepackage{apalike}
\usepackage{combelow}
\usepackage{SCITEPRESS}     

\usepackage{color}
\usepackage[hyphens]{url}

\begin{document}

\title{Wireless Sensor Network based System for the Prevention of Hospital Acquired Infections}

\author{\authorname{Iuliana Bocicor\sup{1}, Maria Dasc\u{a}lu\sup{4}, Agnieszka Gaczowska\sup{2}, Sorin Hostiuc\sup{3}, Alin Moldoveanu\sup{4}, Antonio Molina\sup{5}, Arthur-Jozsef Molnar\sup{1}, Ionu\cb{t} Negoi\sup{3} and Vlad Racovi\cb{t}\u{a}\sup{1}}
\affiliation{\sup{1}SC Info World SRL, Bucharest, Romania}
\affiliation{\sup{2}NZOZ Eskulap, Skierniewice, Poland}
\affiliation{\sup{3}Carol Davila University of Medicine and Pharmacy, Bucharest}
\affiliation{\sup{4}Polytechnic University of Bucharest, Romania}
\affiliation{\sup{5}Innovatec Sensing\&Communication, Alcoi, Spain}
\email{\{iuliana.bocicor,arthur.molnar,vlad.racovita\}@infoworld.ro, maria.dascalu@upb.ro, \{agaczkowska,soraer,negoiionut\}@gmail.com, alin.moldoveanu@cs.pub.ro, amolina@innovatecsc.com}}


\abstract{Hospital acquired infections are a serious threat to the health and well-being of patients and medical staff within clinical units. Many of these infections arise as a consequence of medical personnel that come into contact with contaminated persons, surfaces or equipment and then with patients, without following proper hygiene procedures. In this paper we present our ongoing efforts in the development of a wireless sensor network based cyber-physical system which aims to prevent hospital infections by increasing compliance to established hygiene guidelines. The solution, currently developed under European Union funding integrates a network of sensors for monitoring clinical workflows and ambient conditions, a workflow engine that executes encoded workflow instances and monitoring software that provides real-time information in case of infection risk detection. As a motivating example, we employ the workflow in the general practitioner's office in order to comprehensively present types of sensors and their positioning in the monitored location. Using the information collected by deployed sensors, the system is capable of immediately detecting infection risks and taking action to prevent the spread of infections.}

\onecolumn \maketitle \normalsize \vfill

\section{\uppercase{Introduction}}
\label{sec:introduction}
Hospital acquired infections (HAI) are a serious threat to the health and well-being of both patients and medical staff within clinical units. Various micro-organisms, especially multi drug resistant bacteria may lead to significant hospital related morbidity and mortality. These infections have high related costs and represent a direct occupational hazard for clinical personnel. Hospital infections are a worldwide problem, regardless of geographical, political, social or economic factors \cite{WHO02}, \cite{WHO10}. Furthermore, technological development and sophistication of medical care does not automatically result in lowered infection rates \cite{tikhomirov87}, \cite{coello93}, \cite{WHO10}, \cite{ecdc15}. According to the findings of the World Health Organization, average HAI prevalence in Europe is 7.1\%, in the United States it is 4.5\%. In low- and middle-income countries infection rates vary between 5.7\% and 19.1\% \cite{WHO11}. In intensive care units located in high-income countries the proportion of infected patients can be as high as 51\%, while in low- and middle-income countries it can reach 88.9\% \cite{WHO11}. Unfortunately, many of these infections lead to patient deaths. Annually, infections are accountable for 37 000 deaths in Europe and 99 000 deaths in the USA \cite{WHO11}. While measures and precautions are being taken to successfully reduce these rates \cite{cdc2016}, there is still much room for improvement.

Many hospital infections arise as a consequence of medical personnel that come into contact with contaminated surfaces or equipment, relatives coming in contact with patients or as auto infections, which occur by touching sensitive body parts, such as the face, with the hands. The most common target sites for hospital infection are the urinary and respiratory tracts, which are often involved in minimally-invasive procedures such as catheter-related procedures. 

While the methodology for prevention exists, it is often ignored due to lack of time, unavailability of appropriate equipment or because of inadequate staff training. Research shows that the most important transmission route are staff members who come into contact with patients or contaminated equipment without following proper hygiene procedures \cite{hammer}. More particularly, most often transmission is made by touching a patient or contaminated equipment and then touching another patient without proper hand hygiene \cite{pittet01}. Recent guidelines, such as ''Five moments for hand-hygiene'' \cite{WHO15} provide concise and well-structured information on efficient disinfection means to significantly reduce the risk of infection.

In this paper we present our progress in the development of a cyber-physical system based on a wireless sensor network (WSN) that is targeted towards HAI prevention. Our solution employs a sensor network that will monitor clinical workflows and ambient conditions, integrated with configurable software to detect deviation from established hygiene practices. The sensor network collects information in real-time about substance and material presence. Availability of antimicrobial agents and sterile gloves, as well as environmental conditions that affect pathogen spread, such as oxygen level, airflow, and temperature will be monitored. To complete the picture, the system will facilitate monitoring of complex processes such as management of indwelling urinary catheters, postoperative care, intubation and mechanical ventilation. Given the complexity of these processes, as well as the diversity of hospital regulations, these processes will be described using software workflows. Each clinical process will be modelled using one workflow instance executed by a software workflow engine. When the sequence of transitions inferred by the system from sensor data presents deviations from the expected flow, the system will alert responsible personnel.

The current stage of development represents a system proof of concept, including multifunctional smart sensors for monitoring the use of soap, antimicrobial gel and water sink together with a first clinical workflow that describes the required hygiene procedures in the general practitioner's office. This paper details the smart devices employed, the hardware-software integration as well as the software components that ensure the cyber-physical system achieves its objective of lowering the number and severity of hospital infections.

The present paper is structured as follows. The following section presents some of the last decade's advancements, from an Information and Communication Technologies (ICT) perspective. This includes several software-based or cyber-physical systems developed to increase compliance to guidelines and decrease infection rates. Section \ref{workflow} offers a detailed description of the workflow in the general practitioner's office, as well as the main challenges faced when encoding this process using a software-based workflow model. The wireless sensor network custom designed for the system is depicted in Section \ref{sensor}, where all aspects related to its hardware and software components are covered. Section \ref{system} focuses on presenting the software workflow engine that interprets events and monitors execution of configured workflows. Moreover, the means by which the hardware part of the system is integrated with the software workflow engine, as well as the communication infrastructure employed are illustrated in this section. Finally, we outline our conclusions and further work in the last section.

\subsection{Existing Solutions}
Because HAI have significant detrimental effects, in recent decades health care facilities have started to implement prevention programmes for patients and medical staff. There are even practical guides devised by specialised agencies \cite{WHO12}, which can be used as starting points for the development of good practice plans concerning workplace and patient safety. Following the current technological developments in all medical areas, technology is also present for monitoring and prevention of HAI. This is illustrated by the development of various software-based or cyber-physical solutions that monitor and ensure compliance. In this section, we present some of the most popular such systems.

\subsubsection{Monitoring Hand Hygiene}
Inadequate hand hygiene is responsible for a large proportion of infections \cite{pittet01}. There are several automated solutions to reduce infections caused by improper hand hygiene, most of which use continuous surveillance and immediate notification in case non-compliance is detected \cite{shhedi15}. IntelligentM \cite{intelligentM} and Hyginex \cite{hyginex} are two solutions that monitor employees using bracelet-like devices equipped with Radio Frequency Identification (RFID) technology and motion sensors. Whenever a hygiene event has been omitted, the device alerts them either using vibration (IntelligentM) or coloured lights (Hyginex). Biovigil technology \cite{biovigil} and MedSense \cite{medsense} are designed having the same purpose, only in these cases bracelets are replaced with badges worn by  healthcare workers. The Biovigil device uses chemical sensors to detect whether hand hygiene is observed according to established standards. The system can be configured to remind clinicians to disinfect their hands before entering patient wards, or before administering procedures such as intravenous drips or catheter insertion. Furthermore, these systems record hygiene events, centralise them and enable analysis, visualisation and report generation. SwipeSense \cite{swipe_sense} employs small, alcohol-based devices and wearable gel dispensers. This allows medical personnel to perform hand hygiene without interrupting their activities to go to a sink or disinfectant dispenser \cite{simonette13}. In opposition to the systems mentioned so far, which use sensors placed at patient ward entrances, UltraClenz's Patient Safeguard System \cite{ultraclenz} is ''bed-centric'' and prompts workers to sanitize before and after every patient contact. The DebMed system \cite{debmed} does not use RFID technology, nor any devices for the medical personnel, but instead estimates the number of hand hygiene opportunities per patient-day and compares this number with the actual hand hygiene events that were performed, which are determined using a network of wireless-enabled dispensers.

\subsubsection{Disinfection Robots}
Dangerous pathogens can remain in the air or on different types of surfaces in a hospital room for long periods after the infection source was removed. To tackle this issue, which cannot always be resolved using traditional cleaning and disinfection procedures, several types of disinfection robots have been developed. Generally, they are able to perform thorough disinfection using ultraviolet (UV) light or chemical substances. The Xenex \emph{''Germ-Zapping Robot''} \cite{xenex} can disinfect a room using pulses of high-intensity, high-energy ultraviolet light. The robot must be taken inside the room to be disinfected and in most cases, the deactivation of pathogens takes place in five minutes. Tru-D Smart UVC \cite{trudi} scans the room to be disinfected using eight sensors and computes the optimal short wavelength ultraviolet light dose required for disinfection according to the size, geometry, surface reflectivity as well as the amount and location of equipment found in the room. The robot performs disinfection of the entire room, from top to bottom in one cycle and from one location, ensuring that the ultraviolet light reaches even shadowed areas. The Bioquell Q-10 robots \cite{bioquell} emit a powerful antibacterial bleaching agent, called hydrogen peroxide to kill multi-drug resistant organisms. As hydrogen peroxide is toxic to humans, after disinfection the Q-10 uses another solution to ensure that it is safe for humans to enter the room.

\subsubsection{Managing Infection and Outbreaks}
A different procedure in the fight against infection is implemented by the Protocol Watch decision support system for prevention and management of sepsis \cite{protocol_watch}. Protocol Watch regularly checks certain medical parameters of patients, to reduce the time elapsed between the moment sepsis is first detected and beginning of treatment. If the system detects that certain conditions indicative of sepsis are met, it alerts medical staff and indicates which tests, observations and interventions must be performed, according to established prevention and treatment protocols.

Another goal pursued by clinicians when dealing with hospital infection is the identification of control policies and optimal treatment in infection outbreaks. A comprehensive approach that uses electronic health records to build healthcare worker contact networks is described in \cite{curtis13}. Its main goal concerns putting efficient vaccination policies into place in case of infection outbreaks. 

Among other relevant software systems developed to enhance treatment policy in case of infection outbreak or epidemics are RL6:Infection \cite{rl6} and Accreditrack \cite{accreditrack}. RL6:Infection is a software solution developed to assist hospitals in the processes of controlling and monitoring infections and outbreaks, while Accreditrack is designed to ensure compliance with hand hygiene guidelines, verify  infection management processes as well as to provide procedural visibility and transparency.

\subsection{The HAI-OPS platform}
The proposed platform is developed within the Hospital Acquired Infection and Outbreak Prevention System (HAI-OPS) research project \cite{hai-ops}. Its main objective is to decrease overall mortality and morbidity associated with hospital infection. It is designed to handle both singular infection cases as well as outbreaks, by targeting most common sources and transmission pathways. Operationally, the platform will leverage advances in computing power and availability of custom-developed, affordable hardware that will be combined with a configurable, workflow-based software system \cite{haiops2016}.

Existing solutions, such as those detailed in the section above \cite{intelligentM,hyginex,biovigil,medsense} can be successfully employed to monitor a single process, such as hand hygiene, or equipment and room disinfection \cite{xenex,trudi,bioquell}. While these processes are important for keeping patients and staff safe from infection, there are many other processes that can lead to hospital infection. Among the most prevalent, we mention catheter management, mechanical ventilation, invasive procedures and surgical site care \cite{WHO02,coello93}. One solution for monitoring multiple processes would be to deploy several such systems in parallel. However, given that eHealth interoperability is currently an open issue, this is not only cost-ineffective, but technologically infeasible. We believe that monitoring several clinical and maintenance workflows can be successfully addressed using a single system. Such a system must be configurable so that it covers differences between clinical unit location and layout, differences in types and specifics of undertaken procedures, as well as variation between hygiene guidelines that must be observed by staff. The HAI-OPS platform is designed to address these issues in both hardware as well as software. First of all, using customized, but affordable hardware allows sensors to be deployed in key locations in cost effective manner. Workflow engines allow researchers to create custom BPMN-encoded \cite{bpmn} workflows that encode key events in monitored processes. Furthermore, implementation of a user interface for workflow management will allow epidemiologists to further customize the monitored workflows. To the best of our knowledge, our proposed system is the first of its kind to combine a sensor network and software in a cyber-physical system of the proposed versatility.

\section{\uppercase{General Practitioner's Office Workflow}} 
\label{workflow}
The cyber-physical system depicted in this paper employs pre-defined workflows that describe the processes that the system will monitor. They allow the system to take real-time action in case an infection risk is detected. Our development approach is bottom-up, and starts with modelling some of the less complex workflows, which involve only medical staff and patients. The more complex workflows, that also involve equipment, such as endoscopic or surgical procedures will be addressed at a later time. Thus, the first workflow we approach for the system prototype, which is also the subject of the present paper's motivating example, is the workflow of the general practitioner's office.

\subsection{Process Description}
The general practitioner (GP) is a medical doctor whose practice is not limited to a certain speciality and who provides treatment and preventive care to patients. As opposed to physicians working with inpatients admitted to hospital for certain procedures, the general practitioner works with outpatients, who require consultation or treatments which do not necessitate hospital admission.
All information regarding the GP office, as well as the consultation workflow described were supplied by NZOZ ESKULAP \cite{nzoz}, an outpatient clinic from Poland that is targeted for the first pilot deployment of our system. Figure \ref{Fig:GPOffice} illustrates the general practitioner's office layout from the Polish clinic. The office contains a desk for the physician, a consultation bed and, most importantly for our use case, an area with several elements for ensuring hygienic conditions: a sink, a waste bin and an area dedicated to disinfectants and disinfectant dispensers. The same figure also depicts the planned layout of the wireless sensor network used for monitoring the workflow. These are described in more detail in Section \ref{sensor}.

In order to ensure compliance with the infection-prevention guidelines in the Polish clinic, the first step was identifying the hygiene practices to which the general practitioner must adhere to before, during and after patient consultation. The conventional workflow for an outpatient consultation, including all required actions for ensuring conformity with hygiene standards are depicted within the following sequence of steps:

\begin{enumerate}
	\item Patient enters the office.
	\item The GP starts a conversation with the patient, in order to learn about their medical history, current treatment and reason for the visit. Generally, the physician uses pen and paper or a hospital information system to record information to the patient file.
	\item The GP prepares to examine the patient. The preparation process is crucial with regards to infection prevention. According to current regulations within the target clinic, the doctor must sanitize their hands according to 10 steps for effective hygiene. These are:
	\begin{enumerate}
		\item Wet hands thoroughly.
		\item Soap up, using the liquid soap dispenser. The used tap must be elbow or wrist operated. The physician must rub palms.
		\item Rub palms with finger interlaced.
		\item Massage between fingers, right palm over the left hand and then vice-versa.
		\item Scrub with fingers locked, including fingertips.
		\item Rub rotationally, with thumbs locked.
		\item Rinse thoroughly.
		\item Dry hands using a paper towel that must be placed in proximity to the hand washing facility.
		\item Work towel between fingers.
		\item Dry around and under the nails.
	\end{enumerate}
	\item The GP throws the wet towel to a special waste bin.
	\item The GP starts patient examination.
	\item After the examination, the GP uses an alcohol-based sanitizer for hand disinfection.
	\item The GP goes back to the desk and records examination results using pen and paper or the hospital information system.
	\item Patient leaves the office. 
\end{enumerate}

\begin{figure*}[!t]
	\centering
    	\includegraphics[width=1\textwidth]{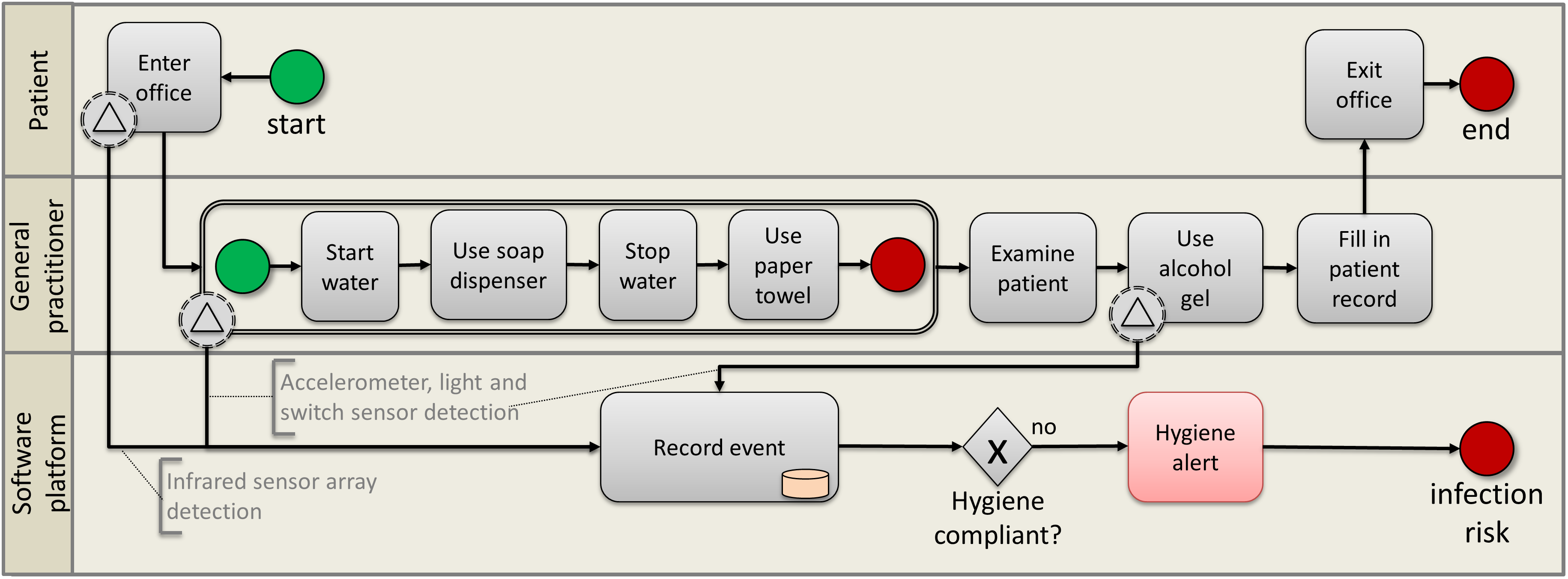}
  	\caption{General Practitioner Workflow}
  	\label{fig:GPWorkflow}
\end{figure*}

The procedure described above concerns a regular examination. However, for special cases such as examinations involving the head, eyes, ears, nose and throat (HEENT), or when the patient presents with skin infection, the doctor must also employ nitrile or latex disposable gloves. Gloves should also be worn whenever there might be contact with blood, body fluids, mucous membranes or non-intact skin. Gloves must be put on immediately before the task to be performed, and removed and discarded as soon as the procedure is completed.

The BPMN workflow for the consultation process is illustrated in Figure \ref{fig:GPWorkflow}. Sections \ref{sensor} and \ref{system} describe how the wireless sensor network is used for monitoring and how the workflow engine monitors the execution of hygiene-relevant events. In the case a deviation from the expected steps of the workflow is detected, a real-time alert is generated and sent to the GP using a mobile device in their possession.

\subsection{Workflow Description}
The general practitioner workflow is shown using BPMN specification in Figure \ref{fig:GPWorkflow}. For the description of the workflow we use both Figures \ref{fig:GPWorkflow} and \ref{Fig:GPOffice}, as the events specified in the workflow are detected by hardware devices placed in different locations in the office. As soon as the patient enters the office, this is detected by the infrared array sensor element placed near the entrance \cite{ilight2016}. The system records and interprets the received data and a workflow instance is started. As illustrated in Figure \ref{fig:GPWorkflow}, the first steps required from the GP is to start the water sink, use the soap dispenser and then stop the sink. The system interprets this as hand hygiene being performed. These events are detected by the sensor elements in the sink and those in the disinfectant dispenser area, which are all connected to smart nodes placed near the physician's desk. This enables transmitting the data to the software server via wireless network. While current regulations described in the previous section require a specific sequence of actions to be undertaken for hand hygiene to be considered effective, our system only checks that the sink and disinfectants were operated. The main reason for this is that the system is envisaged as an additional aid for medical personnel that ensures their safety from possible infection. The system is designed on the principle that medical personnel are responsible and aware of the detailed actions they must undertake to ensure their own, as well as their patients' safety.

The intermediate step of the workflow concerning patient examination starts when the system has detected that hand hygiene compliance is achieved. Otherwise, the system generates and stores a hygiene alert, which is immediately sent to the general practitioner. The two activities are exclusive: if an alert is generated, the workflow instance is stopped and the recorded hygiene breach is recorded. In case of an alert, the GP must perform hand hygiene, after which the system initiates a new workflow instance. In case initial hand hygiene and patient consultation are carried out according to the workflow, the GP must disinfect their hands using antimicrobial gel after the last contact with the patient. This event is again recorded by the system using the same sensors situated in the disinfectant dispenser area and the smart nodes near the physician's desk. The workflow is thus completed. All the information related to patient entry/exit, hygiene compliance and alerts is saved to persistent storage for further reuse, including statistics and advanced analyses for finding the source or propagation of an outbreak.

\subsection{Current Challenges}
Although seemingly straightforward, the process described above can become quite complicated, mainly due to various types of constraints and interferences that may occur. Below we present the main challenges to the system, with regard to the GP office workflow and the methods we use to approach and overcome some of them. Others are still open to discussion and solutions are currently being investigated.

First and foremost, one key aspect to consider is achieving minimal overhead on the clinical process and minimal intrusive interaction, from the user experience point of view. It is important that the system does not impose any constraints and does not restrict the doctor's movements. In many clinical units, including the one targeted for pilot deployment, hospital regulations specify that personnel are not allowed to wear jewellery, watches or bracelets, as these can hamper their freedom of movement and spread bacteria, especially if these wearables are difficult to disinfect. To tackle this, the proposed system does not require the use of additional wearables. Monitoring is done using the deployed wireless sensor network nodes, which are placed in key locations within the GP office, as described in Section \ref{sensor}. In addition, medical personnel already employ chest-mounted badges to which radio-frequency tags can be easily added.

The placement of the wireless sensors is an essential challenge in itself, as locations must be chosen in a manner that allows a complete and preferably optimal surveillance of monitored workflows. The arrangement of sensors in the office must be adjusted to the process, but should also be sufficiently general in order to allow monitoring several workflows: in this case, both the regular consultation workflow as well as HEENT examinations. Thus, in addition to placing wireless sensors at the office entrance, sink, soap or disinfectant dispenser, in order to ensure complete process monitoring, a device is also placed on the waste bin, to detect when gloves are thrown away. Device positioning in the GP office is discussed in more detail within Section \ref{sensor}.

One of the remaining challenges for cyber-physical systems such as the proposed one concerns short-term human interactions that are difficult to detect. In the case of the GP workflow, how should the system detect and react to a person entering the GP office during an examination? In this case, the hygiene event performed by the GP before patient examination is considered cancelled, as the third person can contaminate the physician or patient with micro-organisms. Medical staff wear badges that can be used to identify them using the sensors deployed near the entrance; however, if the person is not part of the medical staff, they cannot be identified. A potential solution is that once the system detects someone entering the office, regardless of whether the person is medical staff or not, the system triggers the execution of a new workflow, including the necessary hygiene events. In case this is not performed, the GP is alerted to take immediate corrective action. 

\section{\uppercase{The Sensor Network}} \label{sensor}
A wireless sensor network consists of a group of electronic devices in which every node controls one or more sensors that measure physical phenomena such as light, heat or proper acceleration. All collected measurements are sent using a wireless network protocol to another device featuring more powerful processing capabilities. Depending on their functionality, nodes are classified into dummy and smart nodes. As the name suggests, dummy nodes consist of small devices that have to effectuate just one simple task: detect a generated event and pass the information to a smart node. A dummy node is particularly characterized by its small size (35x35mm) and low power consumption. Some sensors, like RFID readers, do not generate events by themselves and require a pre-processing stage, which can be exclusively carried out by a powered device. Smart nodes must be able to collect key actions detected from dummy nodes and generate more complex events comprising information regarding four relative clauses: \emph{who} is the person involved, \emph{what} was the action generated, \emph{when} it happened and \emph{where} it happened.

\subsection{Sensor Types}
Required sensors were selected to enable monitoring the clinical workflow detailed in Section \ref{workflow}. From the mentioned steps, sensors in dummy nodes should be applied mainly to detect key actions, such as the utilization of hygienic elements that can be found in the GPs office: water sink, soap dispenser, waste bin, alcohol sanitizer and glove dispenser. The main sensor types required to ensure effective monitoring of the general practitioner office workflow are as follows:

\begin{enumerate}
	\item \textbf{Accelerometer}. These sensors measure changes in gravitational acceleration on two or three axes, allowing to detect changes in motion and orientation. Accelerometers may be attached to water taps, which regulate water flow on the vertical axis and temperature on the horizontal. They may also be applied to sanitizer or glove dispensers, where detected motion implies that they have been used by a practitioner or checked by cleaning staff.         
	\item \textbf{Proximity and light sensor}. Proximity sensors emit infrared radiation and look for changes in the return signal. This type of sensors are already applied in some water sinks and soap containers, but to the best of our knowledge none of them have communication capabilities to report actions.          
	\item \textbf{Switch detection}. A switch is just an electronic component that interrupts the flow of electric current from one conductor to another. It may be operated by a moving object, which makes it a great choice for applications such as detecting the use of waste bins or opening of a door. This is the least energy consuming element from the list, because it does not have to expend energy doing continuous measurement.
\end{enumerate}

Dummy nodes generate action events indicating, for instance, that someone used the soap dispenser or the waste bin, but it is the smart node who has to fill in  information and identify who generated the action. To achieve that goal, it is necessary to process and combine the output from the following two sensors:

\begin{enumerate}
	\item \textbf{Infrared array sensor}. It is a thermopyle type infrared sensor which detects the amount of infrared rays. It has a built-in lens with a 60 degree viewing angle. The sensor offers output for thermal presence, direction and temperature values.  
	\item \textbf{RFID reader}. Radio-frequency identification works using tag-based identification. Tags are small devices similar to stickers that may be carried by people, animals or objects. They can also be easily attached to wearables such as badges or mobile equipment. The frequency range and applied antenna depend on the application and indirectly on the distance between readers and tags. In some clinics, medical and cleaning staff are used to carry a badge with an identification card based on this principle.
\end{enumerate}   

\begin{figure}
	\centering
    \includegraphics[width=0.4\textwidth]{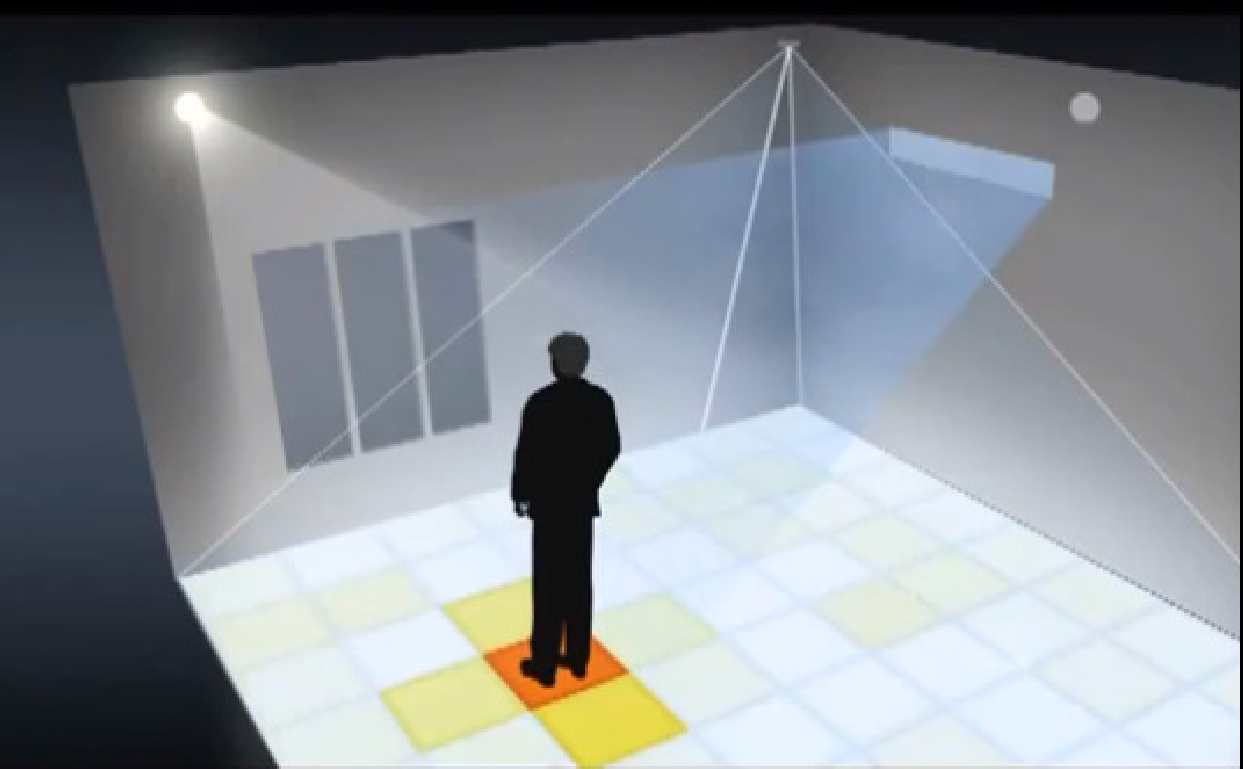}
  	\caption{Infrared array sensor detecting people inside the GP office}
  	\label{GridEye}
\end{figure}

\subsection{Device Positioning}
\begin{figure}
	\centering
    \includegraphics[width=0.4\textwidth]{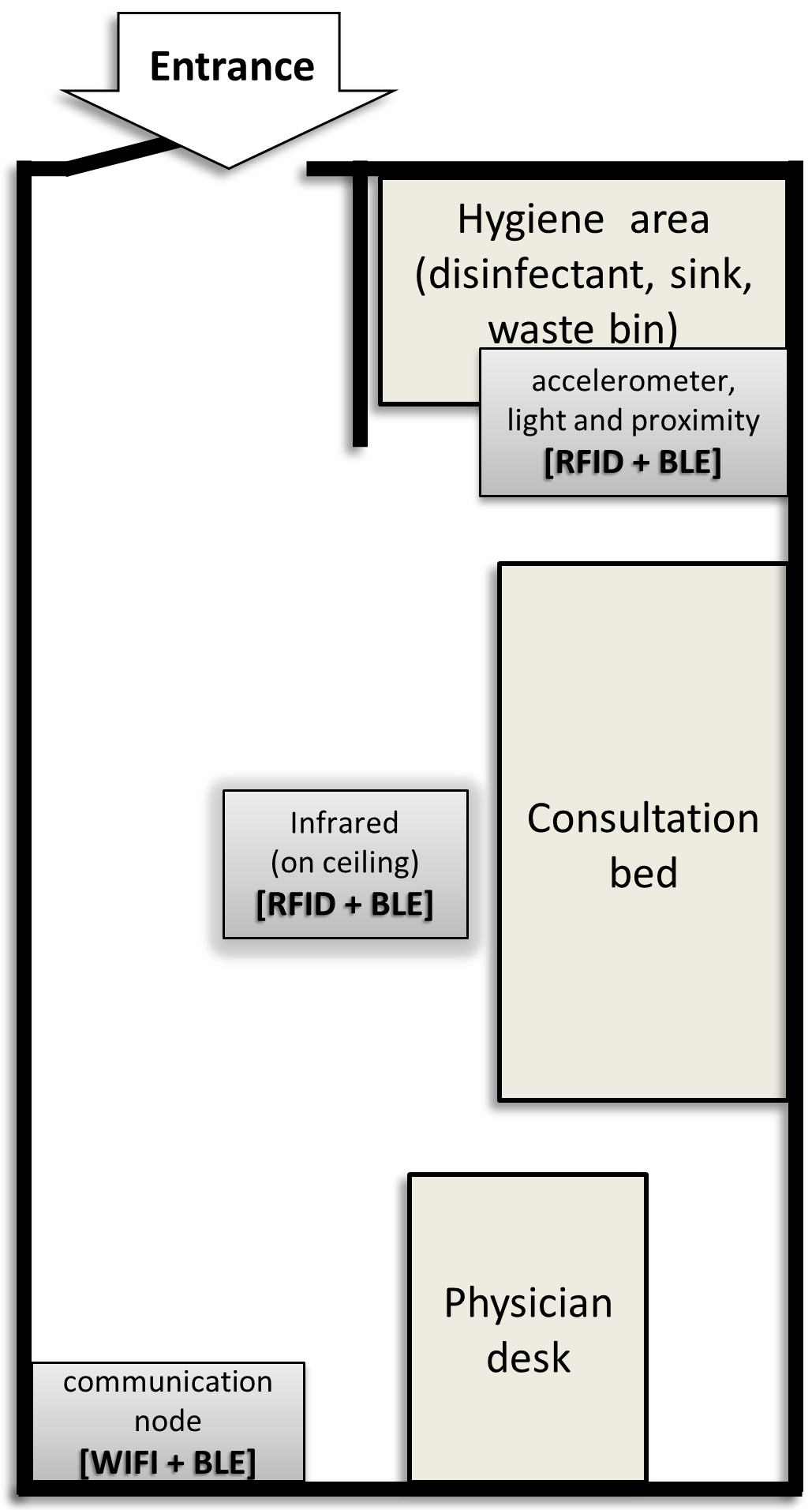}
  	\caption{Layout of the general practitioner office augmented with wireless sensor network}
  	\label{Fig:GPOffice}
\end{figure}

Device positioning and calibration are crucial for the proper functioning of the system. In the case of dummy sensors, the proximity sensor may detect false positives if the distance range is not correctly adjusted or if the sensor is incorrectly placed. When applied to water sinks or gel dispensers, the proximity sensor must be tied to the tap pointing downwards. The system registers when someone places their hand under the tap and when they stop using it. Figure \ref{Fig:GPOffice} illustrates the positioning of both smart and dummy nodes within the general practitioner's office.
 
In addition to the proximity sensor, the RFID antenna and passive infrared array sensor must also be placed according to their detection range. RFID readers provide received signal strength indication (RSSI) levels for detected tags, a measure which is proportional to the distance between them. Patch antennas consist of a planar dielectric substrate material with a radiating patch on one side and a ground plane on the other. The radiating side must point to the GP's office where elements to be identified are located and the ground plane must point to the corridor, ceiling or to an adjacent room. Radio frequency power output must be configured to meet European Union regulations and to avoid false positive detections as much as possible.

Passive infrared array sensors complement the information from RFID readers. If this information is combined properly, the dummy node is able to locate people inside the room, identify people wearing an RFID card (typically clinical staff) and detect people who are not wearing tags (typically patients). The passive infrared sensor must be placed on the ceiling, pointing downwards and centred in the room, as seen in Figure \ref{GridEye}. If the sensor's angle of view is not enough and doesn't fit the whole room, scalability is achieved using several sensors.

\subsection{Communication Protocols}
As already stated, the main features of dummy nodes are their small size and low power consumption. Both features are very closely correlated, because in most cases product size is determined by the battery. In this case, the power consumption in dummy nodes is so low, that they can be powered using coin batteries. The reduced power consumption is due to the integration of a Bluetooth Low Energy (BLE) module \cite{ble}. Compared to previous Bluetooth standards, BLE is intended to provide considerably reduced power consumption and lower cost, while maintaining a communication range of up to 150 meters with connected devices.

Smart nodes have more complex processing, communication and thus, higher power requirements than dummy nodes. They are continuously listening for input BLE connections and when data is received, the result is forwarded to a database server for persistent storage and subsequent analyses. Connection with this server is carried out using existent network infrastructure, regardless of whether it is wired or wireless. Every smart device is identified within the network using a unique IP address and has a fixed location inside the building.

\section{\uppercase{Hardware-Software Integration}} \label{system}
The software side of the system implements the client-server paradigm and employs a software server to which an arbitrary number of heterogeneous clients can have simultaneous connection. The server contains components that receive sensor data from the network, a persistence layer that manages sensor readings as well as a workflow engine that executes workflow instances in real-time, generating alerts when sensor readings indicate deviations from expected workflow transitions. The main software components of the system are as follows:
\begin{enumerate}
	\item \textbf{Connected Device Controller.} The connected devices, or smart nodes, constitute the principal hardware component of the platform. They are responsible for monitoring the clinical environment using sensors and sending sensor readings to the software server. By themselves, they cannot decide whether an infection risk is present. Each connected device includes a software controller, a generic software component that runs independently of the software server. Its objectives are to ensure the correct functioning of smart nodes and to send sensor readings to the software server.

	\item \textbf{Data acquisition.} This is the software component that will be responsible with receiving sensor readings. Received data is stored within the persistence layer, from which it is read and used by  other components. This includes advanced analyses components yet to be developed which do not make the object of the present paper. The server adopts a REST architecture \cite{rest} to receive sensor data acquired as presented in Section \ref{sensor}. The data interchange format employed is JavaScript Object Notation. Information contained in files received from sensors includes the event's timestamp, a Uniform Resource Locator that identifies which node generated the information as well as sensor reading values. For instance, a sensor monitoring temperature will transmit a temperature value in degrees Celsius. A sensor monitoring the presence of an individual will transmit a boolean value, according to whether presence was detected. An RFID reader will transmit the RFID tag identifier and the received signal strength indicator value.

	\item \textbf{Workflow engine adapter.} This software component is a fa\c{c}ade to the workflow engine implementation used by the system. Its main purpose is to abstract the particularities of the workflow engine. This allows the system to operate with any major off the shelf workflow engine implementation. This component provides the required features that allow for the creation, update and deletion of clinical workflows monitored by the system. The workflow engine interprets events, such as inputs from deployed sensors (e.g. hand washing detected), and acts upon them according to a predefined process. The actions are  configurable and can vary from saving a new entry into a database, sending an e-mail or emitting a real-time notification via an external application or short message service. Its input is represented by process descriptions. Processes are composed of activities connected with transitions. Processes represent an execution flow. Each execution of a process definition is called a process instance. As an example, hand disinfection in the general practitioner's office can be represented as a process. Each time a patient enters the office a new process instance is started, managed by the business process management system. Some activities, such as recording an event, or sending an alert are automatic. Others involve waiting for an external event to occur, such as a sensing device reporting the physician has disinfected hands. The workflow engine keeps track of the state of process executions and manages creation and progress of process executions.

	\item \textbf{Data store.} This component of the software server acts as the system's persistence layer for all system data. In addition to user, alert, connected devices and workflow data, it includes a complete record of data received from connected sensors. This is required in order to facilitate advanced analyses for outbreak prevention, identification and monitoring. The data store will be implemented using an SQL database.
\end{enumerate}

\section{\uppercase{Conclusions}}
\label{sec:conclusion}
This paper details the ongoing effort in the development of a cyber-physical system intended for preventing hospital infections. The HAI-OPS platform will integrate a wireless sensor network and a software server that uses a workflow execution engine to monitor key steps within various clinical processes that were identified as responsible for a large proportion of hospital infection. When key steps to ensure process hygiene are not taken, the system will generate real-time alerts. The present paper focuses on describing and modelling an initial clinical process used as motivating example: outpatient consultations within the general practitioner's office. This process is the one selected for implementation during the system's first pilot deployment within a Polish outpatient clinic  \cite{nzoz}. Although we directed our attention specifically towards the motivating workflow, the system is designed to allow for deployment of diverse sensor network configurations, as well as facilitate creation and execution of many different clinical workflows.

Upcoming system development will build on existing achievements. First of all, the system will be used to model more complex clinical workflows, including endoscopic and minor surgery procedures, which will be implemented in the pilot site location. Second of all, as part of the project a graphical component will be developed to allow management of monitored workflows. In addition, we aim to leverage available sensor readings by implementing advanced reporting and evaluation capabilities. These are expected to help clinical epidemiologists in pinpointing infection and outbreak sources using visualizations such as risk maps and healthcare worker contact networks \cite{Hladish2012}.

\section*{\uppercase{Acknowledgement}}
This work was undertaken as part of the HAI-OPS project funded by the European Union, under the Eurostars programme\footnote{https://www.eurostars-eureka.eu/project/id/9831}.

\vfill
\bibliographystyle{apalike}
{\small
\bibliography{bibliography}}

\begin{thebibliography}{}

\bibitem[{Bioquell}, 2016]{bioquell}
{Bioquell} (2016).
\newblock Bioquell q-10.
\newblock
  \url{http://www.bioquell.com/en-uk/products/life-science-products/archive-hc-products/bioquell-q10/}.

\bibitem[{BIOVIGIL Healthcare Systems, Inc.}, 2015]{biovigil}
{BIOVIGIL Healthcare Systems, Inc.} (2015).
\newblock Biovigil and our team.
\newblock \url{http://www.biovigilsystems.com/about/}.

\bibitem[{{B}luetooth {SIG}, {I}nc.}, 2017]{ble}
{{B}luetooth {SIG}, {I}nc.} (2017).
\newblock Bluetooth low energy.
\newblock
  \url{https://www.bluetooth.com/what-is-bluetooth-technology/how-it-works/low-energy}.

\bibitem[Bocicor et~al., 2016]{haiops2016}
Bocicor, M.~I., Molnar, A.-J., and Taslitchi, C. (2016).
\newblock Preventing hospital acquired infections through a workflow-based
  cyber-physical system.
\newblock In {\em Proceedings of the 11th International Conference on
  Evaluation of Novel Software Approaches to Software Engineering}, pages
  63--68.

\bibitem[{Centers for Disease Control and Prevention}, 2016]{cdc2016}
{Centers for Disease Control and Prevention} (2016).
\newblock {HAI} data and statistics.
\newblock \url{https://www.cdc.gov/hai/surveillance/}.

\bibitem[Coello et~al., 1993]{coello93}
Coello, R., Glenister, H., Fereres, J., Bartlett, C., Leigh, D., Sedgwick, J.,
  and Cooke, E. (1993).
\newblock The cost of infection in surgical patients: a case-control study.
\newblock {\em Journal of Hospital Infections}, 25:239--250.

\bibitem[Curtis et~al., 2013]{curtis13}
Curtis, D., Hlady, C., Kanade, G., Pemmaraju, S., Polgreen, P., and Segre, A.
  (2013).
\newblock Healthcare worker contact networks and the prevention of
  hospital-acquired infections.
\newblock {\em Plos One}.
\newblock DOI: 10.1371/journal.pone.0079906.

\bibitem[{DebMed - The Hand Hygiene Compliance and Skin Care Experts},
  2016]{debmed}
{DebMed - The Hand Hygiene Compliance and Skin Care Experts} (2016).
\newblock A different approach to hand hygiene compliance.
\newblock
  \url{http://debmed.com/products/electronic-hand-hygiene-compliance-monitoring/a-different-approach/}.

\bibitem[{European Centre for Disease Prevention and Control}, 2015]{ecdc15}
{European Centre for Disease Prevention and Control} (2015).
\newblock Annual epidemiological report. antimicrobial resistance and
  healthcare-associated infections. 2014.
\newblock
  \url{http://ecdc.europa.eu/en/publications/Publications/antimicrobial-resistance-annual-epidemiological-report.pdf}.

\bibitem[{Excelion Technology Inc.}, 2013]{accreditrack}
{Excelion Technology Inc.} (2013).
\newblock Accreditrack.
\newblock \url{http://www.exceliontech.com/accreditrack.html}.

\bibitem[Fielding, 2000]{rest}
Fielding, R.~T. (2000).
\newblock Architectural styles and the design of network-based software
  architectures.
\newblock Doctoral dissertation, University of California.

\bibitem[{General Sensing}, 2014]{medsense}
{General Sensing} (2014).
\newblock Medsense clear. hand hygiene compliance monitoring.
\newblock \url{http://www.generalsensing.com/medsenseclear/}.

\bibitem[Goga et~al., 2016]{ilight2016}
Goga, N., Vasilateanu, A., Mihailescu, M.~N., Guta, L., Molnar, A.-J., Bocicor,
  I., Bolea, L., and Stoica, D. (2016).
\newblock Evaluating indoor localization using wifi for patient tracking.
\newblock In {\em International Symposium on Fundamentals of Electrical
  Engineering (ISFEE)}.

\bibitem[{HAI-OPS}, 2017]{hai-ops}
{HAI-OPS} (2017).
\newblock home page.
\newblock \url{http://haiops.eu}.

\bibitem[Hammer, 2013]{hammer}
Hammer, S. (2013).
\newblock Hand washing: Reducing nosocomial infections.
\newblock
  \url{http://iwsp.human.cornell.edu/files/2013/09/Hand-Washing-Reducing-Nosocomial-Infections-2j1mlfb.pdf}.

\bibitem[Hladish et~al., 2012]{Hladish2012}
Hladish, T., Melamud, E., Barrera, L.~A., Galvani, A., and Meyers, L.~A.
  (2012).
\newblock Epifire: An open source c++ library and application for contact
  network epidemiology.
\newblock {\em BMC Bioinformatics}, 13(1):76.

\bibitem[{Hyginex}, 2015]{hyginex}
{Hyginex} (2015).
\newblock Introducing hyginex generation 3.
\newblock \url{http://www.hyginex.com/}.

\bibitem[{NZOZ Eskulap}, 2016]{nzoz}
{NZOZ Eskulap} (2016).
\newblock {NZOZ} eskulap.
\newblock \url{www.eskulapskierniewice.pl/}.

\bibitem[{Object Management Group}, 2015]{bpmn}
{Object Management Group} (2015).
\newblock Business process model and notation.
\newblock \url{http://www.bpmn.org/}.

\bibitem[{Philips}, 2015]{protocol_watch}
{Philips} (2015).
\newblock Protocolwatch - {SSC Sepsis}.
\newblock
  \url{http://www.healthcare.philips.com/main/products/patient_monitoring/products/protocol_watch/}.

\bibitem[Pittet, 2001]{pittet01}
Pittet, D. (2001).
\newblock Improving adherence to hand hygiene practice: a multidisciplinary
  approach.
\newblock {\em Emerging Infectious Diseases}, 7:234--240.

\bibitem[{RL Solutions}, 2015]{rl6}
{RL Solutions} (2015).
\newblock The rl6 suite / infection surveillance.
\newblock \url{http://www.rlsolutions.com/rl-products/infection-surveillance}.

\bibitem[Ryan, 2013]{intelligentM}
Ryan, J. (2013).
\newblock Medtech profiles: Intelligentm - a simple yet powerful app to
  dramatically reduce hospital-acquired infections.
\newblock \url{https://medtechboston.medstro.com/profiles-intelligentm/}.

\bibitem[Shhedi et~al., 2015]{shhedi15}
Shhedi, Z.~A., Moldoveanu, A., Moldoveanu, F., and Taslitchi, C. (2015).
\newblock Real-time hand hygiene monitoring system for hai prevention.
\newblock In {\em The 5th IEEE International Conference on E-Health and
  Bioengineering - EHB 2015}.

\bibitem[Simonette, 2013]{simonette13}
Simonette, M. (2013).
\newblock Tech solutions to hospital acquired infections.
\newblock
  \url{http://www.healthbizdecoded.com/2013/06/tech-solutions-to-hospital-acquired-infections/}.

\bibitem[{Swipe Sense}, 2015]{swipe_sense}
{Swipe Sense} (2015).
\newblock Hand hygiene. redefined.
\newblock \url{https://www.swipesense.com/}.

\bibitem[Tikhomirov, 1987]{tikhomirov87}
Tikhomirov, E. (1987).
\newblock Who programme for the control of hospital infections.
\newblock {\em Chemioterapia}, 6:148--151.

\bibitem[{Tru-D Smart UVC}, 2016]{trudi}
{Tru-D Smart UVC} (2016).
\newblock About tru-d.
\newblock \url{http://tru-d.com/why-uvc-disinfection/}.

\bibitem[{UltraClenz}, 2016]{ultraclenz}
{UltraClenz} (2016).
\newblock Patient safeguard system.
\newblock \url{http://www.ultraclenz.com/patient-safeguard-system/}.

\bibitem[{World Health Organization}, 2002]{WHO02}
{World Health Organization} (2002).
\newblock Prevention of hospital-acquired infections - a practical guide.
\newblock
  \url{http://www.who.int/csr/resources/publications/whocdscsreph200212.pdf}.

\bibitem[{World Health Organization}, 2010]{WHO10}
{World Health Organization} (2010).
\newblock The burden of health care-associated infection worldwide.
\newblock \url{http://www.who.int/gpsc/country_work/summary_20100430_en.pdf}.

\bibitem[{World Health Organization}, 2011]{WHO11}
{World Health Organization} (2011).
\newblock Health care-associated infections - fact sheet.
\newblock
  \url{http://www.who.int/gpsc/country_work/gpsc_ccisc_fact_sheet_en.pdf}.

\bibitem[{World Health Organization}, 2012]{WHO12}
{World Health Organization} (2012).
\newblock Prevention of hospital-acquired infections - a practical guide.
\newblock \url{http://apps.who.int/medicinedocs/documents/s16355e/s16355e.pdf}.

\bibitem[{World Health Organization}, 2015]{WHO15}
{World Health Organization} (2015).
\newblock Clean care is safer care - five moments for hand hygiene.
\newblock \url{http://www.who.int/gpsc/tools/Five_moments/en/}.

\bibitem[{Xenex}, 2015]{xenex}
{Xenex} (2015).
\newblock Xenex germ-zapping robots.
\newblock \url{http://www.xenex.com/}.

\end{thebibliography}

\vfill

\end{document}